\documentclass[twoside]{article}
\usepackage{fleqn,epsfig,espcrc2,float}
\usepackage{graphicx}
\usepackage[figuresright]{rotating}

\newcommand{\AmS}{{\protect\the\textfont2
  A\kern-.1667em\lower.5ex\hbox{M}\kern-.125emS}}
\title{ Towards a complete $\gamma\gamma \rightarrow 4~$leptons Monte Carlo }

\author{C. Carimalo, W. da Silva, F. Kapusta \\
\vspace{1pc}
Laboratoire de Physique Nucl\'eaire et de Hautes Energies,  \\
IN2P3-CNRS Universit\'es Paris VI et VII, \\
4, Place Jussieu - BP200, Tour 33 - Rez de Chauss\'ee, 
75252 PARIS Cedex 05, FRANCE   }
       
\begin{document}

\begin{abstract}
The total cross-section of the $\gamma \gamma \rightarrow 4 $ leptons 
process is known to have a large constant value at high energy, 
but a complete and exact calculation of related differential spectra 
does not  exist yet. However a complete study of the process is important 
at NLC for luminosity measurements or for background estimations. 
We here report on our various steps in that direction, including 
the implementation of a first new calculation of helicity amplitudes 
in a Monte-Carlo generator.
 
\vspace{1pc}
\end{abstract}

\maketitle

\section{MOTIVATIONS}

The high-energy limit of the total cross-section of the $\gamma \gamma 
\rightarrow 4 $ leptons process was computed 30 years ago and found 
to be large \cite{LiFr,Serb,ChWu} : 
$$\sigma(\gamma \gamma \rightarrow e^+e^-e^+e^-) \quad =\quad 6500  
\quad \mbox{nb} $$
$$ \sigma(\gamma \gamma \rightarrow  e^+e^- \mu^-\mu^+)\quad =
\quad 5.7 \quad \mbox{nb} \quad  $$
$$\sigma(\gamma \gamma \rightarrow \mu^-\mu^+\mu^-\mu^+)\quad 
=\quad 0.16\quad \mbox{nb} $$
This property is important enough for considering 
those processes as ``reference processes'' for luminosity measurements 
at high-energy $\gamma \gamma $ colliders. 
As expected, the bulk of the cross-section is provided by 
kinematical regions where the outgoing lepton-pairs are close to the 
$\gamma \gamma $ beam axis. This is clear from the paper by Kuraev et 
al.\cite{Kur} where they computed single lepton angular spectrum at small angle 
in the ``logarithm approximation''. 

It is worthwhile to check whether $\gamma \gamma \rightarrow e^+ e^-\ X$ 
and/or $\gamma \gamma \rightarrow \mu^-\mu^+\ X$ could compete with  
single lepton-pair production which is the basic process for luminosity
measurement. This requires an exact computation at relatively large angles.

Besides, $\gamma\gamma$ into two lepton-pairs might be a background to 
rare processes, as for example  $J / \psi$ production. 
In the same spirit, hybrid pairs production should be considered in    
$\gamma \gamma \rightarrow \tau^+ \tau^-$ studies. 

An exact computation of $\gamma \gamma$ double lepton-pair production
is also useful for heavy ions peripheral collisions and TESLA.

Other interesting topics in astrophysics are the interactions of a  
high energy cosmic $\gamma$ with background infrared (IR) $\gamma$,  
and with the earth atmosphere.

The $e^+e^-$ single and double pair production from $\gamma\gamma_{IR}$ 
collisions have cross sections of the same order of magnitude around 
a squared invariant mass s$_{\gamma\gamma}$ of about 0.6 GeV$^2$, 
obtained for example with 4 PeV $\gamma$ scattering on 35$\mu m$ wavelength 
photons. Their interaction with 2.7 K background ones is of course 
the dominant one. However this constant cross section should not be forgotten 
for ultrahigh energy photons.\\
 With  $\sigma_2  = \sigma ( \gamma \gamma  \rightarrow
 e^+ e^- e^+ e^- )$ asymptotic value,
 the ratio of  
 $$\sigma( \gamma Z  \rightarrow Z e^+ e^- e^+ e^- ) \simeq 
 \sigma_2 Z^2 \frac{\alpha}{\pi} \ln \frac{2\omega}{m_e}$$  
 and 
 $$\sigma( \gamma Z  \rightarrow Z e^+ e^- ) \simeq 
 \frac{28}{9} Z^2 \alpha \frac{\alpha^2}{m_e^2} \ln \frac{2\omega}{m_e}$$
 is of the order of  $\sigma_2 / \sigma_{\mathrm{Thomson}}  \simeq 10^{-5}$
 and rather negligible in shower development. 

\section{EXACT QED COMPUTATION} 
 
\begin{figure}[htb]
   \begin{center}
          \epsfxsize=0.45\textwidth
          \epsffile{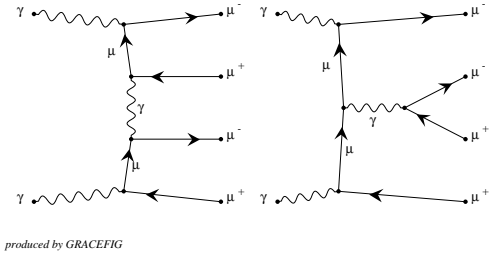}
   \end{center}
\caption{Basic diagrams with space-like and time-like exchanged $\gamma$.}
\label{fig-diag}
\end{figure}

 64 amplitudes are to be computed, as the basic ones shown 
 in figure~\ref{fig-diag}, including permutation and  
anti-symmetrisation for $\gamma \gamma  \rightarrow
 \mu^+ \mu^- \mu^+ \mu^- $,  where the exchanged $\gamma$ is either space-like 
or time-like. At large angles, masses can be neglected and we are left with 12 
irreducible helicity amplitudes which are expressed in terms 
of two elementary functions $S=\bar{U^{ \prime} \uparrow} U \downarrow$
and $T= \bar{U^{\prime}  \downarrow } U\uparrow $, \`a la Kleiss-Stirling.
The  $d \sigma(\gamma \gamma  \rightarrow
 \mu^+ \mu^- \mu^+ \mu^- )$ is then folded in a Monte Carlo generator to 
 get some realistic distributions. In addition, to avoid numerical singularities 
 which might arise for massless final particles, masses 
 have been included in the propagators. 

For cross-checking purposes, two factorization approaches have been 
used and they are described in the following.
\section{FACTORIZATION FORMULAS}
\begin{figure}[ht]
   \begin{center}
          \epsfxsize=0.4\textwidth
          \epsffile{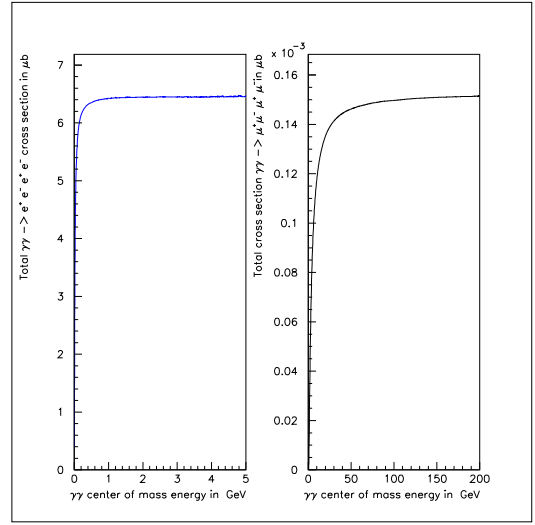}
   \end{center}
\caption{Energy dependence of the cross-sections $\sigma(\gamma\gamma \rightarrow e^+e^-e^+e^-$)  
and $\sigma(\gamma \gamma \rightarrow \mu^+ \mu^-\mu^+ \mu^- $).}
\label{fig-sig}
\end{figure}

\begin{figure}[H]
   \begin{center}
          \epsfxsize=0.35\textwidth
          \epsffile{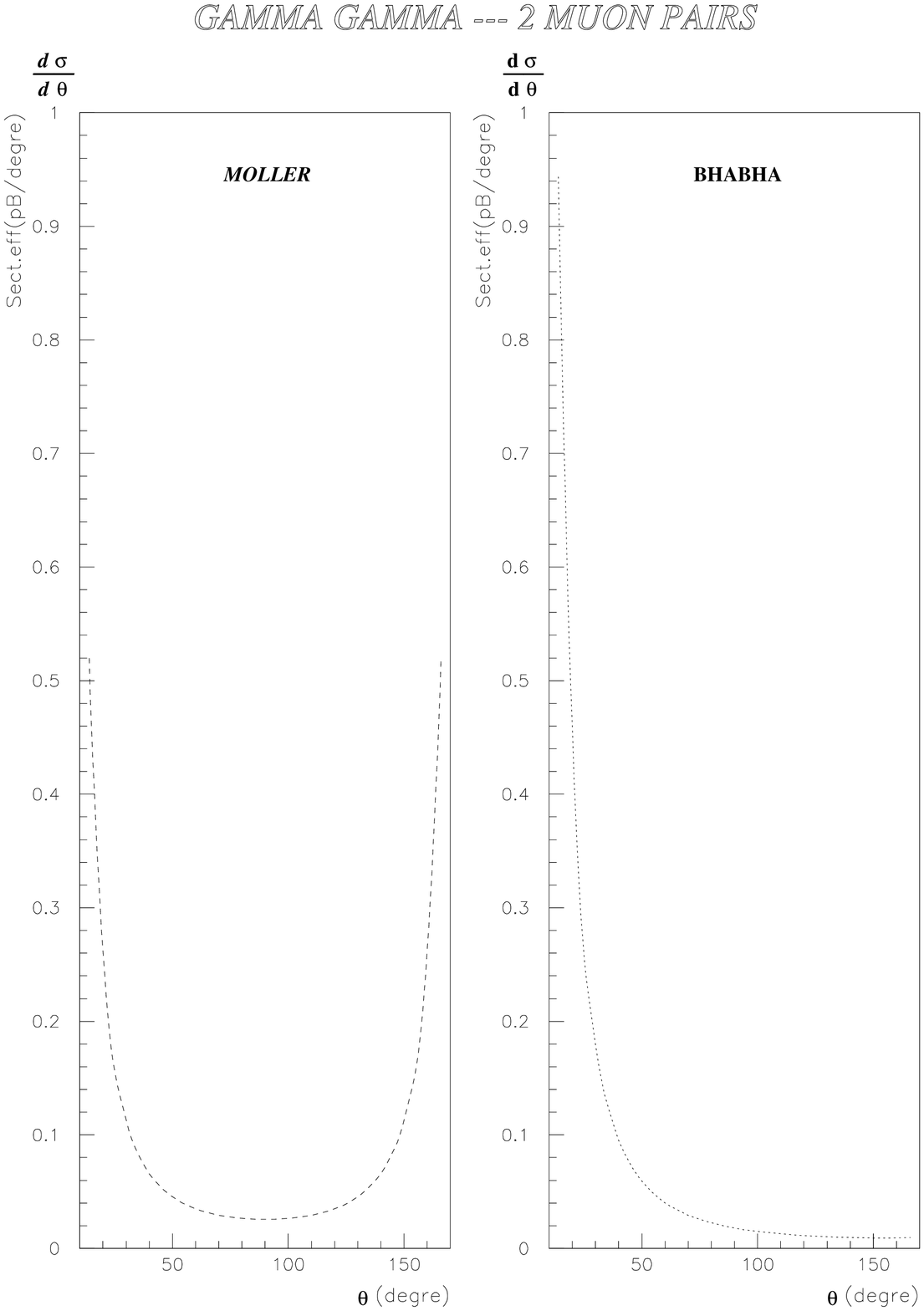}
   \end{center}
\caption{$\mu$ angular distributions.} 
\label{fig-muon}
   \begin{center}
          \epsfxsize=0.35\textwidth
          \epsffile{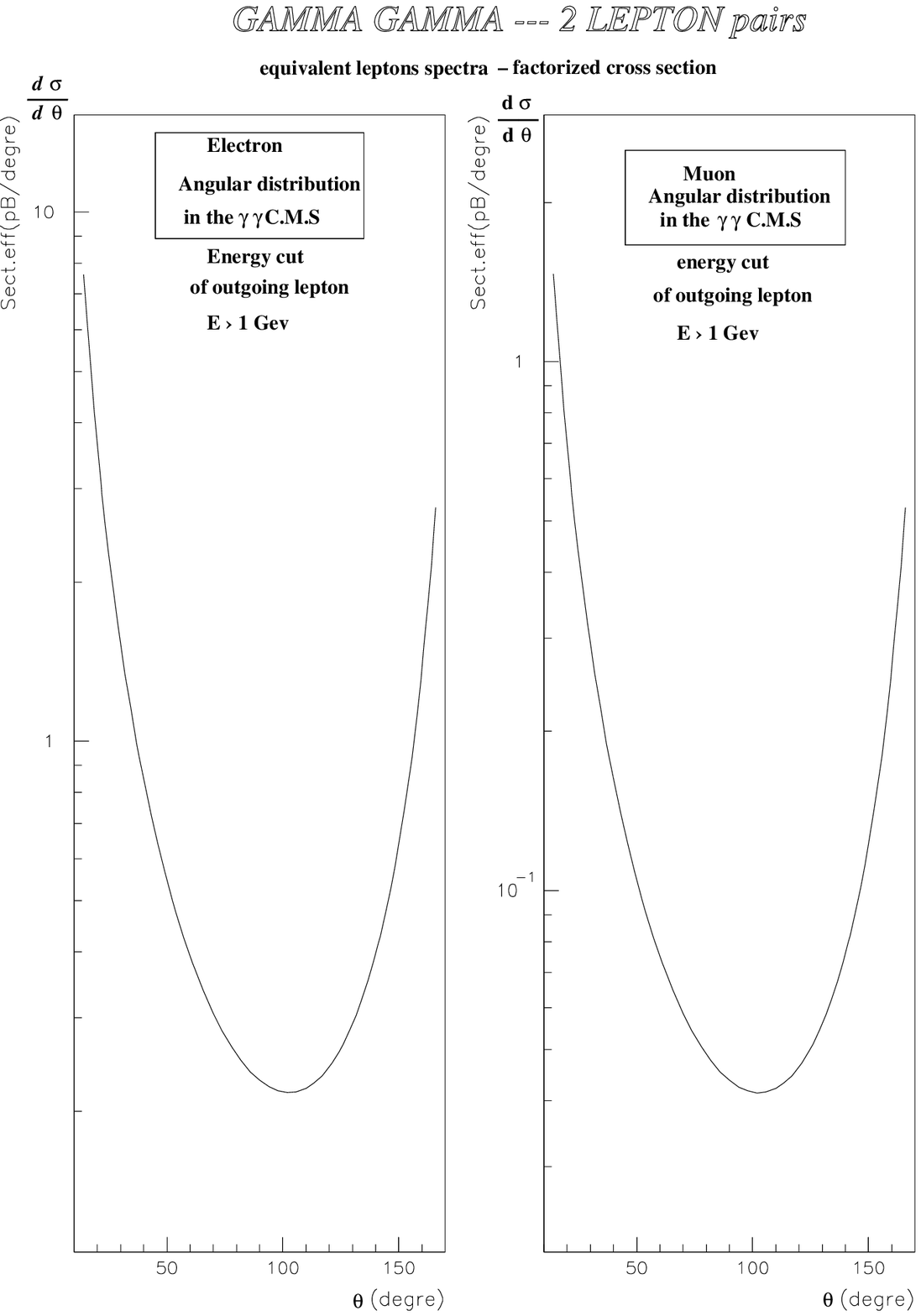}
   \end{center}
\caption{$e$ and $\mu$
angular distributions.}
\label{fig-lept}
\end{figure}

 We have checked that the factorization formula from Carimalo et al. 
 \cite{Car}, gives by analytical integration 
 the well-known high energy limit total cross section :  
$$\sigma(\gamma\gamma \rightarrow e^+e^-e^+e^-) = \frac{\alpha^4}{{m_e}^2}
 \frac{1}{\pi}(\frac{175}{36}\zeta(3)-\frac{19}{18})$$
 The lepton-pair invariant mass spectra have been also derived. 
  A numerical integration allows one to obtain the energy dependence of total 
  cross-sections. Figure~\ref{fig-sig} shows clearly for the first time the threshold 
 effect for  $\gamma\gamma \rightarrow e^+e^-e^+e^-$  
and $\gamma \gamma \rightarrow \mu^+ \mu^-\mu^+ \mu^- $.
As a further check, a factorized lepton spectrum has been used 
to describe the contribution of lepton-lepton scattering subprocess 
\cite{Cohen}. The angular distributions are shown in figures~\ref{fig-muon}
and~\ref{fig-lept}. 
At  $\sqrt{s_{\gamma\gamma}} = 400~$GeV, 
with two leptons of energy greater than 1 GeV above 10$^\circ$ 
and the two remaining ones below 2$^\circ$,   
the visible cross section is $\simeq$  0.4 pb for muons 
and 2 pb for electrons.  
For comparison purposes,
 $\sigma(\gamma \gamma \rightarrow \mu^+ \mu^-) = 25$ pb  and 
 $\sigma(\gamma \gamma \rightarrow e^+ e^-) = 42.5$ pb .
With an acceptance cut of 5$^\circ$ and 10$^\circ$ these cross-sections 
are lowered respectively to 8.6 pb and 6 pb.
With 5 nb$^{-1}$s$^{-1}$, a few $\%$ precision on the luminosity measurement
is reached within 1 day. A priori, 2 lepton-pair production does not improve much. 
 However at lower angles, further studies are needed.
\section{OUTLOOK}
\begin{itemize}
\item Complete the computation (time-like photon graphs).
\item Put masses on final state particles.
\item Include the EW contribution.
\item Compare with GRACE...
\item Give realistic numbers for various configurations.
\end{itemize}

\end{document}